\documentclass{llncs}

\usepackage{makeidx}  
\usepackage{color}
\usepackage{amssymb}
\usepackage{amsmath}
\usepackage{dsfont}
\usepackage{graphicx}

\newcommand{\upshift}[0]{\vspace{-0.5em}}

\newcommand{\XX}{\mathcal{X}}
\newcommand{\XC}{\mathcal{C}}
\newcommand{\XD}{\mathcal{D}}
\newcommand{\XO}{\mathcal{O}}

\newcommand{\relbb}{\stackrel{\scriptscriptstyle{H}}{\sim}}
\newcommand{\relsc}{\stackrel{\scriptscriptstyle{H}}{\approx}}
\newcommand{\XZ}{\mathds{Z}}

\title{Equivalence Classes of Optimal Structures in HP~Protein Models
Including Side Chains}

\titlerunning{Equivalence Classes in HP~models}

\author{
	Martin Mann
	\and Rolf Backofen
	\and Sebastian Will
}
\authorrunning{Mann et al.}

\institute{
	\vspace{-0.5em}
	University of Freiburg, Bioinformatics, 79110 Freiburg, Germany,\\
	\email{\{mmann,backofen,will\}@informatik.uni-freiburg.de}
}

\begin{document}

\maketitle


\begin{abstract}

Lattice protein models, as the Hydrophobic-Polar (HP) mo\-del, are a common
abstraction to enable exhaustive studies on structure, function, or evolution of
proteins. A main issue is the high number of optimal structures, resulting from
the hydrophobicity-based energy function applied.
We introduce an equivalence relation on protein structures that correlates to
the energy function. We discuss the efficient enumeration of optimal
representatives of the corresponding equivalence classes and the application of
the results.

\end{abstract}

\section{Introduction}

Proteins are the central players in the game of life. They are involved in
almost all processes in cells and organisms, comprising replication, metabolism,
and movement. To be able to perform their specific functions, proteins have to
adopt a certain fold or structure, which is encoded by the protein's sequence.
Thus, knowledge of a protein's structure elucidates the mechanisms it is
involved.

Currently, it is not possible to calculate a protein's functional fold from
its sequence nor to simulate the whole folding process in detail. Simplified protein
models are used to reduce the computational complexity. A common abstraction are
lattice proteins~\cite{Dill:95a,Bromberg:94}. Here, the structure space a
protein can adopt is discretized and allows for efficient folding
simulations~\cite{Mann_LatPack_HFSP_08,Steinhoefel:07}. Nevertheless, it is
difficult to determine minimal energy structures, which represent the functional
folds in such models. Even in the most simple Hydrophobic-Polar
(HP)~model~\cite{Lau:89a}, the optimal structure prediction problem stays
computationally hard (NP-complete)~\cite{Berger:98}. Despite this complexity, a
fast calculation of non-symmetrical optimal structures in the HP~model is
possible using constraint programming techniques applied in the
\emph{Constraint-based Protein Structure Prediction (CPSP)}
approach~\cite{Backofen:06a,Mann_CPSPweb_2009,Mann:08a,Will:06}.

Recently, we have introduced a significantly improved local search scheme for
lattice protein folding simulations~\cite{Ullah:09} using a full
Miyazawa-Jernigan energy potential~\cite{Miyazawa:MJ:96}. We take advantage of
the efficient CPSP approach and initialize the folding simulations with optimal
structures from the simpler HP~model. This incorporates the phenomenon of
hydrophobic collapse of protein structures, a driving force at the beginning of
the folding process~\cite{Agashe:95}. The already compact structures from the
CPSP application form the starting point of the folding driven by more complex
interactions. This scheme outperforms folding simulations using a standard
initialization with random structures and yields better results within shorter
simulation time~\cite{Ullah:09}.
To increase efficiency of local search methods, usually many optimization runs
from different starting points are done.
Since the set of all HP-optimal structures is usually too large as a starting
set, we are interested in a smaller subset that still covers the structural
diversity of the whole set as good as possible. We achieve this by enumerating
optimal structures that maintain a given minimal distance to each other.

Due to the hydrophobicity-focusing energy function, proteins in HP~models show
on average a huge number of optimal structures. Since polar residues do not
contribute to the energy, optimal structures usually show a much higher
variation in the placement of polar than hydrophobic residues.

Here, we introduce an equivalence relation to partition the set of (optimal)
structures into according classes. Two structures are defined to be equivalent,
iff they do not differ in the placements of their hydrophobic residues. We
introduce an extension to the CPSP approach that enables an efficient
calculation of the number of equivalence classes of optimal structures via
enumerating one representative per class. The approach is presented for
backbone-only and side chain incorporating HP~models. We show that a sequence's
number of representatives (later defined as core-degeneracy) is several
magnitudes smaller than the overall number of all optimal structures
(degeneracy).

Thus, the set of optimal representatives is well placed to be used within the
combined approach of CPSP and local search~\cite{Ullah:09}. Furthermore, we
propose another application of the equivalence classes: Since the equivalence
relation is highly correlated to the HP~energy function, the number of classes
might be a better measure of structural stability than a sequences'
degeneracy~\cite{Shortle:92}.

\section{Preliminaries}
\label{sec-prelim}

\begin{figure}[t]
\vspace{-1em}
\begin{center}
	a)
	\begin{minipage}[t]{0.3\textwidth}
		\includegraphics[width=\textwidth]{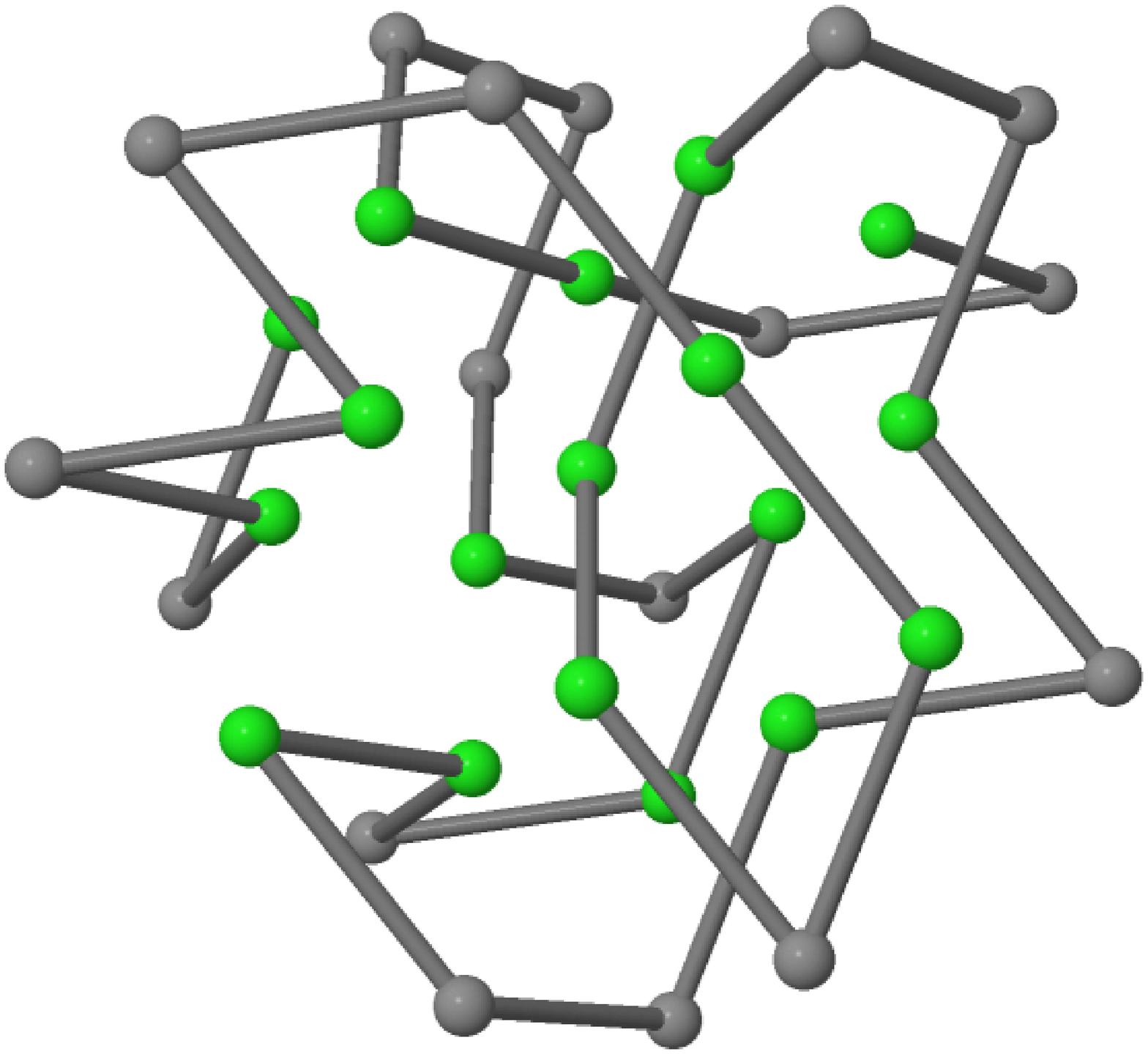}
    \end{minipage}
	\hspace{2em}
	b)
	\begin{minipage}[t]{0.3\textwidth}
		\includegraphics[width=\textwidth]{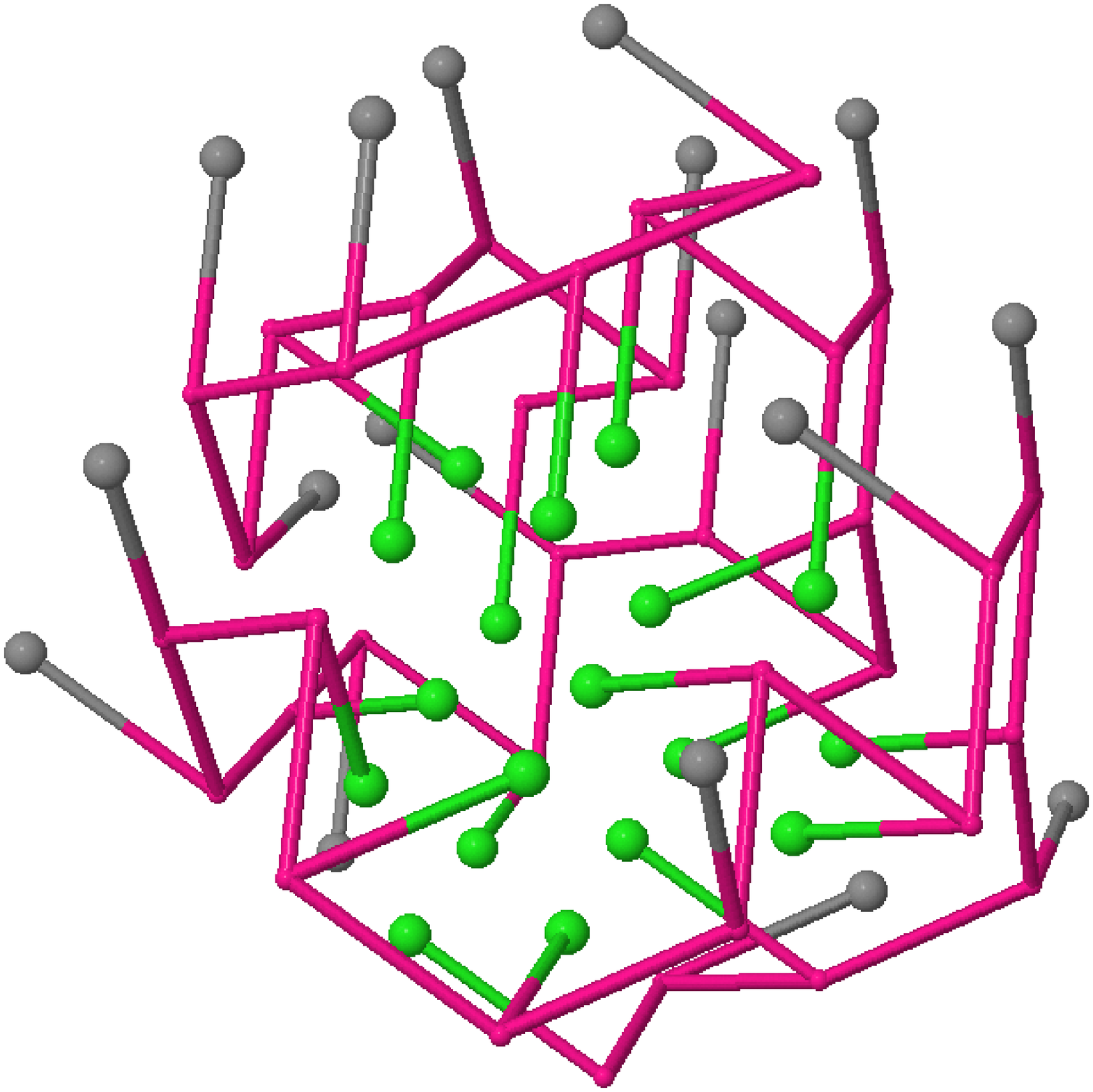}
    \end{minipage}
	\vspace{-0.5em}
	\caption{Optimal structures of
	{\small\texttt{HPPHHPPPHPHHPHHPPHPHPPHHHPHHPPHPHPH}} in the
	face-centered-cubic lattice. (a) backbone-only model with energy~-50,
	(b) side chain model with energy~-55. Colors: green - H~monomers, 
	gray - P~monomers, red - backbone in side chain models. Visualization by
	\texttt{HPview} from CPSP-package~\cite{Mann:08a}.}
	\label{fcc-struct}
\end{center}
\vspace{-2em}
\end{figure}

A lattice protein in the HP~model is specified by its sequence~$S\in\{H,P\}^n$,
where~$H$ and~$P$ denote hydrophobic and polar monomers, respectively. The
structure positions are confined to nodes of a regular lattice~$L \subseteq
\XZ^3$. A valid \emph{backbone-only} structure~$C \in L^n$ of length~$n$ is a
self-avoiding walk (SAW) in the underlying lattice~$L$, i.e. it holds
connectivity $\forall_{1\leq i<n}: (C_i-C_{i+1})\in N_L$ and self-avoidance
$\forall_{1\leq i<j\leq n}:C_i \neq C_j$, where~$N_L$ denotes the set of
distance vectors between neighbored points in~$L$. An example is shown in
Fig.~\ref{fcc-struct}a). The energy of a lattice protein structure is given by
non-consecutive HH-contacts:
\begin{equation}
E(S,C)=\sum_{\stackrel{1\leq i<j\leq n}{(i+1)<j}} 
\begin{cases}
	-1 &: (C_i-C_j)\in N_L \wedge S_i=S_j=H \\  
	\;\;\;0 &: \text{otherwise}
\end{cases}
\label{eq:E_HP}
\end{equation}

An \emph{optimal structure} minimizes the energy function. The number of optimal
structures is denoted as \emph{degeneracy} of a sequence and is an important
measure of structural stability~\cite{Shortle:92}.

The \emph{CPSP-approach} by Backofen and Will~\cite{Backofen:06a} enables the
calculation of a sequence's degeneracy without full structure space
enumeration~\cite{Will:06}. It utilizes the observation that optimal structures
show a (nearly) optimal packing of H~monomers. Thus, the CPSP-approach can be
sketched in two major steps:

\begin{enumerate}
  \item \emph{H-core construction:} Given the number~$n_H$ of H~monomers from
  the target sequence~$S$, all optimal packings of $n_H$~monomers are
  calculated. These \emph{optimal H-cores} show the maximal number
  of contacts possible. For a fixed sequence~$S$ and the corresponding~$n_H$,
  we denote the set  of optimal H-cores with $\XO$. The calculation of~$\XO$
  is computationally difficult on its own and was solved by us using
  constraint programming~\cite{Backofen:01b,Backofen:06a}.

  \item \emph{Structure threading:} Given~$S$ and~$\XO$ only structures are
  enumerated where the H~monomers of~$S$ are confined to an optimal
  H-core~$O\in\XO$, i.e. they are ``threaded'' through the H-cores. Since
  all~$O$ show the maximally possible number of contacts between H~monomers,
  each resulting structure is optimal according to Eq.~\ref{eq:E_HP} as well. The
  structure threading is done by solving a Constraint Satisfaction Problem
  (CSP) for each~$O\in\XO$ as given below.
\end{enumerate}

Since step~1 depends only on the number of H~monomers~$n_H$ and no further
property of any sequence, we can precalculate the H-cores for different~$n_H$
and store them in a database. This significantly speeds up the approach and
reduces the computation time to step~2, i.e. the structure threading.

It might happen, that we find no appropriate structure threading for a
sequence~$S$ and the according set of optimal H-cores~$\XO$. Thus, we revert to
the set of the best suboptimal H-cores~$\XO'$ that show at least one contact
less than an optimal H-core~$O\in\XO$ and iterate the procedure. Still it holds:
the first successive structure threading is an optimal structure, since no
H~monomer packing with more contacts was found before. Further details on
the CPSP approach in~\cite{Backofen:06a}.

The CSPs solved in step~2 are given by $(\XX,\XD,\XC)$, where we denote the set
of variables~$\XX$, their domains~$\XD$, and a set of constraints~$\XC$. For
each monomer~$S_i\in S$ a variable~$X_i\in\XX$ is introduced. The SAW is modeled
by a sequence of binary neighboring constraints
$\operatorname{neigh}(X_i,X_{i+1})$ and a global $\operatorname{alldiff}(\XX)$
to enforce the self-avoidingness. The optimal H-core~$O\in\XO$ is used to define
the domains~$\XD$: $\forall_{i : S_i = H} : \XD(X_i) = O$ and $\forall_{i : S_i
= P} : \XD(X_i) = L\setminus O$. Thus, if we find a solution of such a CSP, i.e.
an assignment $a_i\in\XD(X_i)$ for each variable that satisfies all constraints
in~$\XC$, it will minimize the energy function in Eq.~\ref{eq:E_HP}, i.e. an
optimal structure.

\upshift
\section{Representative Optimal Structures}
\label{sec-approach}

Revisiting the CSP we can see, that P~mo\-no\-mers are constrained only by the
SAW constraints. Imagine a sequence with a long tail of P~monomers. Each valid
placement of the subchain in front of the tail can be combined with a
combinatorial number of possible SAWs of the tail. This leads to the immense
degeneracy in the HP~model.

Therefore, we set up an \emph{equivalence relation~{\footnotesize$\relbb$}} on
structures (Eq.~\ref{eq-equiv}) that decomposes the set of all (optimal)
structures into equivalence classes. In the following, the number of equivalence
classes of optimal structures is denoted as \emph{core-degeneracy}. As given by
Eq.~\ref{eq-equiv}, structures from different equivalence classes differ in at
least one H~monomer placement.
\begin{equation}
	C \relbb \hat{C} \Leftrightarrow \forall_{i|S_i =H} : C_i = \hat{C}_i.
	\label{eq-equiv}
\end{equation}

The representative enumeration (that corresponds to core-degeneracy calculation)
can be done via an extension of the CPSP approach presented in
Sec.~\ref{sec-prelim}. Instead of calculating all optimal structures, we want to
calculate only one representative per equivalence class. This has to be ensured
at two stages: (I) the solutions of each single CSP for a given H-core have to
be different according to~Eq.~\ref{eq-equiv}, and (II) the solutions from two
CSPs for two different H-cores have to be different as well. The second
condition (II) holds by definition, because~{\footnotesize$\relbb$} is only
defined on the H~monomer placements that are constrained by different H-cores
from~$\XO$ (differing in at least one position). In the following, we will
discuss how to achieve the difference for solutions of a single CSP (I).

Note that the core-degeneracy, i.e. the number of different placements of
H-monomers, or core-configurations, in optimal structures of a sequence, is
\emph{not equal} to the number of different H-cores, which are the sets of
lattice points that are occupied by H-monomers. The latter number is easily
obtained from the standard prediction algorithm, described in
Sec.~\ref{sec-prelim}. It equals the number of cores, where the sequence is
successfully threaded on.

\subsection*{Restricted Search for Enumeration of Representatives}

The standard way to solve a CSP is a combination of domain filtering (i.e.
constraint propagation) and depth first search. This results in a binary tree
where each node represents a subproblem of the initial CSP (root) and edges
represent the additional constraints added to derive the two subproblems from
its predecessor node (CSP). The constraints~$c$ and~$\lnot c$ added to derive
the leave nodes of a certain CSP are often of the form $c=(X_i\equiv d)$ by
selecting a variable $X_i$ from $\XX$ and a value $d\in\XD(\XX_i)$ according to
some heuristics. The constraint solver traverse the binary tree until a solution
was found or an inconsistency of a constraint from~$\XC$ was detected.

Therefore, a straightforward way to enumerate only one representative for each
equivalence class can be sketched as follows: first, we restrict the search of
the solving process onto the H~associated variables. Then, we perform a single
check for satisfiability, i.e. search for a single assignment of P~monomer
variables fulfilling all constraints in~$\XC$. Thus, we get only one P~monomer
placement for a given H~monomer assignment if any exists.

The drawback of this approach is that we restrict the variable order of the
search heuristics. But the performance of the CPSP~approach mainly depends on
the search heuristics applied to select a certain variable or value from its
domain. It turned out that a mixed assignment of H~and P~associated variables
yields the best runtimes. These heuristics can not be applied within the
sketched procedure where we have to first assign H-associated variables, then
P-associated ones. Thus, a lower CPSP performance is expected. But, we
have to do less search which results in much faster runtimes than enumerating
all optimal structures.

\upshift
\section{Representative Optimal Structures with Side Chains}

Recently, we have introduced the extension of the CPSP
approach~\cite{Mann_CPSPweb_2009} to \emph{HP~models including side
chains}~\cite{Bromberg:94}. Here, each amino acid of a protein sequence is
represented by two monomers: $C_i^b$ representing the backbone atoms, and
$C_i^s$ representing the atoms of the side chain. Beneath the SAW condition on
the backbone monomers~$C_i^b$, we constrain each side chain to be neighbored to
its backbone, i.e. $\forall_{1\leq i\leq n} : (C_i^s-C_i^b)\in N_L$. An example
structure is given in Fig.~\ref{fcc-struct}b). The applied energy function~$E'$
exploits only HH-contacts of side chain monomers~$C_i^s$:
\begin{equation}
E'(S,C^s)=\sum_{1\leq i<j\leq n} 
\begin{cases}
	-1 &: (C_i^s-C_j^s)\in N_L \wedge S_i=S_j=H \\  
	\;\;\;0 &: \text{otherwise}
\end{cases}
\label{eq:E_HP_sc}
\end{equation}

Therefore, the side chain models show an even higher degeneracy than the
backbone-only models discussed so far, since all backbone monomers~$C_i^b$ are
unconstrained by the energy function as well. Thus, an equivalence
relation~{\footnotesize$\relsc$} that focuses on the monomers constrained by the
energy function is even more striking in HP~models including side chains. The
relation~{\footnotesize$\relsc$} is given by
\begin{equation}
	(C^b,C^s) \relsc (\hat{C}^b,\hat{C}^s) 
	\Leftrightarrow
	\forall_{i : S_i = H} : C_i^s = \hat{C}_i^s
\end{equation}  

Therefore, we will enforce that structures from one equivalence class show the
same H~monomer side chain positioning. The CPSP approach for HP~models including
side chains differs only in the CSP formulation from the original approach for
backbone-only models~\cite{Mann_CPSPweb_2009}. This allows for the application
of the same approach discussed in the previous section to enumerate
non-equivalent optimal structure representatives. Thus, we restrict search to
the H~associated side chain variables first and only check for satisfiability on
the remaining variables.

\upshift
\section{Results and Discussion} \label{sec-result}

We exemplify the enumeration of representatives for backbone-only and side
chain models. We focus on the comparison of the resulting core-degeneracy of a
sequence and its overall number of optimal structures, i.e. degeneracy, because
we are interested in a reduced set of optimal structures, e.g. for local
search initialization (see introduction).
All following results are given for HP-sequences of length~27 in 3D~cubic
lattice. Since the enumeration and check of all $2^{27}$~sequences ($>10^8$) is
computationally not feasible, we restrict each study to a large randomly chosen
subset of~$10^5$ and $10^4$~sequences, respectively.

\newcommand{\HPrep}[0]{\textsc{HPrep}}
\newcommand{\XB}{\mathcal{B}}
\newcommand{\XS}{\mathcal{S}}

The program \HPrep{} implements the approach from section~\ref{sec-approach}.
It is integrated into the CPSP-tools package~\cite{Mann:08a} version 2.4.0 and
available online\footnote{\texttt{http://cpsp.informatik.uni-freiburg.de}}.

As discussed in Sec.~\ref{sec-prelim}, the structure threading step of the CPSP
approach screens through a precomputed list of appropriate H-cores in decreasing
number of contacts stored in a database. Therefore, it might occur that the
available list from the database is exceeded without any solution, i.e. no
optimal structure was computed. Still, the energy of the last H-core tried is a
close lower bound on the energy this sequence can adopt. In the following,
$\XB$~denotes the subset of sequences where the current H-core database is not
sufficient and thus the CPSP approach can give only a lower bound for now. The
number of sequences in~$\XB$ is quite small. It is reasonable to assume that the
degeneracy distribution among~$\XB$ is the same as for the remaining sequences
or on average even higher.

\vspace{1em}
\noindent{\bfseries Backbone-only models}\\
\vspace{-0.6em}

\begin{figure}[t]
\vspace{-1em}
\begin{center}
	\begin{minipage}[t]{0.55\textwidth}
		\includegraphics[width=\textwidth]{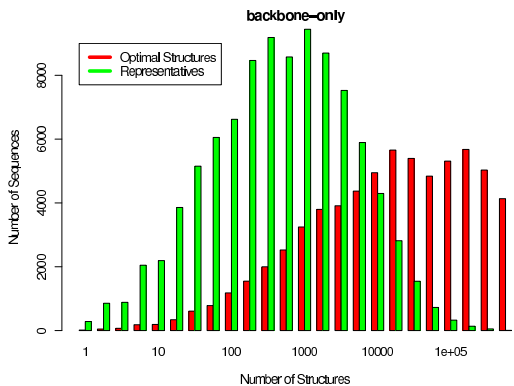}
    \end{minipage}
	\vspace{-0.7em}
	\caption{Backbone-only models : 
	Histogram of core-degeneracy (green) and degeneracy (red) with cut-off~$\leq
	10^6$. (Plots refer to 3D~cubic lattice and sequence length~27.)}
	\label{deg-shift-bb}
\end{center}
\vspace{-2em}
\end{figure}

We tested $10^5$~random sequences in the backbone-only model in the 3D~cubic
lattice. Here, only 66\%~show a degeneracy below~$10^6$. $\XB$ comprises
about~4\% of the sequences. The remaining 30\%~can adopt even more than
$10^6$~structures with minimal energy.

Figure~\ref{deg-shift-bb} summarizes the results: in \emph{red} the degeneracy
and in \emph{green} the core-degeneracy distribution with cut-off~$10^6$ is
presented. Thus, in \emph{red} the degeneracy distribution comprises 66\%~of the
sequences as given above. In contrast, \emph{all} sequences show a number of
optimal equivalence classes below~$10^6$ (in \emph{green})! The average
degeneracy is reduced from~$124800$ (with cutoff~$10^6$) to a mean
core-degeneracy of~$4856$. This reduction within two orders of magnitude results
in reasonably small sets of representative structures e.g. to be utilized in
local search initializations. Furthermore, the enumeration of representatives
is on average six times faster than the enumeration of all optimal structures
with a mean runtime of 2~seconds (Opteron~2356 - 2.3~GHz).

This increase of small sets of representatives compared to the complete sets of
optimal structures shows the advantage of the approach: core-degeneracy does not
show the huge combinatorial explosion of degeneracy. This gets even more
striking in HP~models including side chains, as shown in the next section.

\vspace{1em}
\noindent{\bfseries Models including side chains}\\
\vspace{-0.5em}

\begin{figure}[t]
\vspace{-1em}
\begin{center}
	\includegraphics[width=0.5\textwidth]{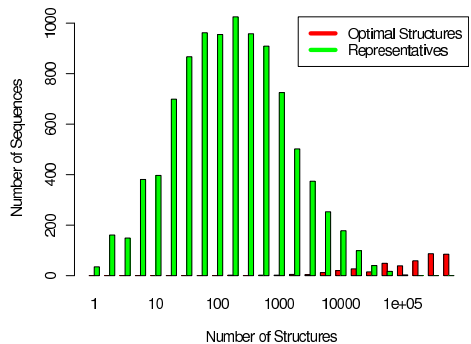}
	\vspace{-1em}
	\caption{Models with side chains : 
	Histogram of the core-degeneracy \emph{(green)} and degeneracy
	\emph{(red)} with cut-off~$\leq 10^6$. Note: only~408 sequences out of~$10^4$
	showed a degeneracy below~$10^6$ as given in the text.
	(Plots refer to 3D~cubic lattice and sequence length~27.)}
	\label{deg-shift-sc}
\end{center}
\vspace{-2em}
\end{figure}

The degeneracy in HP~models including side chains is much higher than for
backbone-only models. This results from the simple energy function
(Eq.~\ref{eq:E_HP_sc}) that does not constrain the backbone or P~monomers.
Therefore, an immense number of optimal structures is present.
From the $10^4$~HP-sequences tested only~408 show a degeneracy below~$10^6$.
$\XB$~comprises again about~3.1\% of the sequences.

When investigating core-degeneracy the picture changes completely: \emph{All} of
the sequences tested have less than~$10^6$ representatives.
Figure~\ref{deg-shift-sc} summarizes the distribution. The average number of
representatives is about~1550, which is again at least three orders of magnitude
smaller than the average degeneracy. Since we have only a very rough lower bound
of~$10^6$ on the average degeneracy (due to the cut-off), the real reduction
ratio is expected to be even higher.

\upshift
\section{Conclusions}

\upshift
The introduced equivalence relations for HP~models enables a energy function
driven partitioning of structures. The presented CPSP approach extension enables
an efficient calculation of representatives for all equivalence classes of
optimal structures, i.e. calculation of a sequence's core-degeneracy. Using
our implementation \HPrep{}, we showed that sequences show several orders of
magnitudes less optimal equivalence classes than optimal structures. This is
most striking in models including side chains.

The sets of representatives are usually small. Furthermore, representatives show
different hydrophobic core arrangements. Therefore, they are well placed to be
used for the initialization of local search procedures that utilize more complex
energy functions~\cite{Ullah:09}. This emulates the hydrophobic collapse in the
folding process.

Since a sequence's degeneracy is a measure of structural
stability~\cite{Shortle:92}, we propose another application of our approach. The
core-degeneracy might be used as a more reasonable \emph{measure of stability}
in the HP~model compared to degeneracy. It ignores the HP~model specific
degeneracy blow-up due to unconstrained subchains of P~monomers (see
section~\ref{sec-approach}). Thus, a structural stability analysis could be
based on the presented equivalence classes instead of all possible structures.

\upshift


\end{document}